\documentclass[article,twocolumn,showpacs,preprintnumbers,amsmath,amssymb]{revtex4}
\usepackage{txfonts}
\usepackage[flushleft]{caption2}
\usepackage{epsfig}
\usepackage{graphics}
\usepackage{overpic}
\begin{document}

\title{Majorana zero modes on a 1D chain for quantum computation}
\author{Lei Chen, W. LiMing$^\dagger$, Jia-Hui Huang}
\address{Department of physics, Laboratory of Quantum Engineering and Quantum Materials,
School of Physics and Telecommunication Engineering, South China Normal University, Guangzhou 510006, China\\
$*$Corresponding author: $^\dagger wliming@scnu.edu.cn$}
\date{\today}

\begin{abstract}
Numerical calculations for Majorana zero modes on a one-dimensional chain are performed using the technique of block diagonalization  for general parameter settings. It is found that Majorana zero modes occur near the ends of the chain and decay exponentially away from the ends. The phase diagrams show that Majorana zero modes of a long-enough chain indeed have a parameter domain of $2t>|\mu|$ as predicted from the bulk property of the chain, but a short chain has a much smaller parameter domain than the prediction. Through a numerical simulation Majorana zero modes are found to be robust under the disturbance of noise. Finally the reversion of the parity of the ground states is studied by applying a bias voltage on the quantum dot at an end of the chain. It is found that for a weak coupling between a chain and a quantum dot the parity of the ground states can be reversed through adiabatically tuning the bias voltage.
\end{abstract}

\maketitle \indent

\section{Introduction}

 In particle physics, Majorana fermions are their own antiparticle\cite{Majorana,Wilczek,Roman,Beenakker}. It is still unclear whether Majorana fermions exist in the nature as elementary building blocks, but they can be constructed in pairs in condensed matter systems\cite{gongming,Sau,Sarma}. 
 Dirac fermion operators $a$ and $a^\dagger$ can be decomposed into a pair of  Majorana operators $\gamma_1$ and $\gamma_2$, e.g. $a=(\gamma_1 + i\gamma_2)/2$. 
 They satisfy $\gamma_i=\gamma_i^\dag$ and an unusual anticommutation relations $\{\gamma_i,\gamma_j\}=2\delta_{ij}$.


One decade ago Kitaev proposed a  basic theoretical model to exhibit Majorana fermions\cite{Kitaev}. On a quantum wire with a superconducting coupling each local Dirac fermion is divided into two local Majorana fermions. Two unpaired Majorana zero modes (i.e. Majorana fermions) appear at the two ends of the wire. 
This model has been extensively studied in literatures up to recent years\cite{Martin,J. Alicea}. The Majorana zero modes give doubly-degenerate ground states of the system,$|\tilde 0\rangle$ and $|\tilde 1\rangle$, with parities $\pm1$, respectively. 

A  superposition of the two degenerate ground states $\alpha|\tilde 0\rangle + \beta|\tilde 1\rangle$
can be used as a qubit, where $\alpha$ and $\beta$ can be expressed by spherical angles in the Bloch sphere. 
Majorana zero modes satisfy the braiding statistics. An interchange of two adjacent Majorana fermions turns $\gamma_1$ and $\gamma_2$ to be $\gamma_2$ and $-\gamma_1$. Alicea {\it et al} proposed a T-junction structure to interchange two Majorana fermions through sequent switches of bias voltages on different parts of the junction\cite{J. Alicea}. 
The interchange, however, can only rotate the initial state by an angle of $\pi/2$ around $z$-axis of the Bloch sphere. 
This indicates that 
a universal quantum gate can not be performed through an interchange of Majorana fermions.

Karsten recently proposed a method for arbitrary rotations to the initial state using a coupling between Majorana zero modes and a quantum dot\cite{Karsten}. By tuning the bias voltage on the quantum dot the parity of the initial state  is reversed. Finally universal quantum gates are realized using four Majorana fermions and three quantum dots with appropriate magnetic flux.

In this paper, we study numerically the Majorana zero modes in Kitaev's model including the wave functions, the phase diagram, the robustness of Majorana zero modes and the parity reversion of the ground states for quantum computation. We use a technique of block diagonalization in Schur's decomposition. It is very different from that of Bogoliubov de Gennes' equations, which does not hold the commutation relations of fermions. We first give a brief introduction to Kitaev's model and the diagonalization method in section II and then report the numerical results of Majorana zero modes, parities and the parity reversion of the ground states in the following three sections. The last section gives a summary.

\section{1D chain}

We begin from a 1D chain of N sites with a superconducting pairing potential. Each site can be either empty or occupied by a spinless electron
. The Hamiltonian reads\cite{Kitaev,A. Kitaev}
\begin{align}
 H_0&=\sum^{N-1}_{j=1}[-t a_j^\dag{a}_{j+1}+\Delta a_ja_{j+1}+h.c.]-\mu \sum^{N}_{j=1}(a_j^\dag{a}_j-\frac{1}{2})
\label{H}
\end{align}
where $a_j^\dag$ is the electron creation operator localized at site j, $t$  the hopping amplitude, $\mu$ the chemical potential, and $\Delta=|\Delta|e^{i\theta}$ the p-wave superconducting gap with a phase $\theta$. One may try to diagonalize the above Hamiltonian using Bogoliubov de Gennes (BdG) equation after a transformation $\hat f_j = \sum_i (u_{ji} a_i + v_{ji} a^\dag_i)$. This method, however, is not valid because the new operators $\hat f_j$ do not satisfy fermion commutation relations, e.g. $\{\hat f_j, \hat f_k \}\ne 0$. 
Kitaev provided a method by introducing the following Majorana fermion operators\cite{Kitaev,A. Kitaev}
\begin{align}
\gamma_{2j-1}=a_{j}e^{i\theta/2}+ a^\dagger_{j}e^{-i\theta/2}\\
 \gamma_{2j}=-i(a_{j}e^{i\theta/2}-a^\dagger_{j}e^{-i\theta/2})
 \end{align}
They satisfy the commutation relation of Majorana fermions $\{\gamma_{i}, \gamma_{j}\} =2\delta_{ij}$ and $\gamma_{i}^\dagger=\gamma_{i}$.  Using these Majorana operators Hamiltonian \eqref{H} is transformed into
\begin{align}
\nonumber H_0&=\frac{i}{2}\sum^{N-1}_{j=1}[(t+|\Delta|)\gamma_{2j}\gamma_{2j+1}+(-t +|\Delta|)\gamma_{2j-1}\gamma_{2j+2}]
\\&-\frac{i}{2}\mu\sum^N_{j=1} \gamma_{2j-1}\gamma_{2j}\equiv \frac{i}{4}\sum_{i=1,j=1}^{2N} \gamma_iA_{ij}\gamma_j
\label{Hchain}
\end{align}
where $A$ is a real skew-symmetric matrix whose nonzero eigenvalues are purely imaginary numbers and come in pairs $\pm i\epsilon_j (\epsilon_j>0) $ for an even-dimension case\cite{Kitaev,A. Kitaev,Alexei}. It allows one to bring A to a block diagonal form through a real orthogonal matrix transformation as follows
\begin{eqnarray}
W A W^T &=& \left(
         \begin{array}{ccccc}
           0 & \epsilon_1  \\
           -\epsilon_1 & 0  \\
           \, & \, & \ddots \\
           \, & \, & \, & 0 & \epsilon_N \\
           \, & \, & \, & -\epsilon_N & 0 \\
         \end{array}
       \right)
       \label{Adia}
\end{eqnarray}
In terms of \eqref{Adia}  Hamiltonian \eqref{Hchain} is reduced into a canonical form
\begin{align}
H_0&=\frac{i}{2}\sum_{m=1}^{N}\epsilon_m \tilde\gamma_{2m-1} \tilde\gamma_{2m}\\&=\sum_{m=1}^N\epsilon_m (\tilde{a}_m^{\dag} \tilde{a}_m-\frac{1}{2}),\quad \epsilon_m\geq0 \label{cano}
\end{align}
where $\tilde\gamma_m=\sum_j W_{mj}\gamma_j$ and $\tilde{a}_m=(\tilde\gamma_{2m-1}+i\tilde\gamma_{2m})/2, \tilde{a}_m^{\dag}=(\tilde\gamma_{2m-1}-i\tilde\gamma_{2m})/2$. Note that $\tilde\gamma_m=\tilde\gamma_m^\dagger$ are Majorana fermion operators and $\tilde{a}_m^{\dag}, \tilde{a}_m$ are Dirac fermion operators representing quasiparticles of the present model. The single-particle energies $\epsilon_m$ have been arranged in an increasing order as $m$ increases. Majorana zero-modes appear when $\epsilon_1$ vanishes. They in general distribute at boundaries of a system. A pair of Majorana zero modes composes a nonlocal Dirac fermion.

\section{Majorana zero modes}

The chain's bulk energy can be obtained through a fourier transformation $a_k = N^{-1/2}\sum_i e^{-ikR_i} a_i$. Up to a constant, the Hamiltonian takes the form
\begin{eqnarray}
H_0= \sum_{k}\phi_k^\dag
\left(
                                                 \begin{array}{cc}
                                                   -2t\cos k-\mu & -2i\Delta\sin k \\
                                                   2i\Delta\sin k & 2t\cos k+\mu \\
                                                 \end{array}
                                               \right)\phi_k
\end{eqnarray}
where $\phi_k^\dag=( a_{k}^\dag, a_{-k})$.  The bulk energy spectra are given by
\begin{eqnarray}
\epsilon_\pm(k) = \pm\sqrt{(2t\cos k+\mu)^2+4\Delta^2\sin^2 k}, \quad |k|\leq\pi.
\end{eqnarray}



The gap between these two energies closes at $2t=|\mu|$. Kitaev provided the zero modes of a special case  $|\Delta|=t>0, \mu=0$\cite{Kitaev,A. Kitaev}. Two unpaired Majorana zero modes with zero energy appear sharply at the two ends of the chain.
Kitaev made a conjecture and a mathematical analysis that Majorana zero modes exist in the domain $|\mu|<2t$. It is believed that Majorana zero modes in a general case distribute near the ends and decay exponentially away from the ends. We first verify this conjecture through numerical computations.

\begin{figure}
\begin{overpic}[width=2.8cm,height=2.8cm]{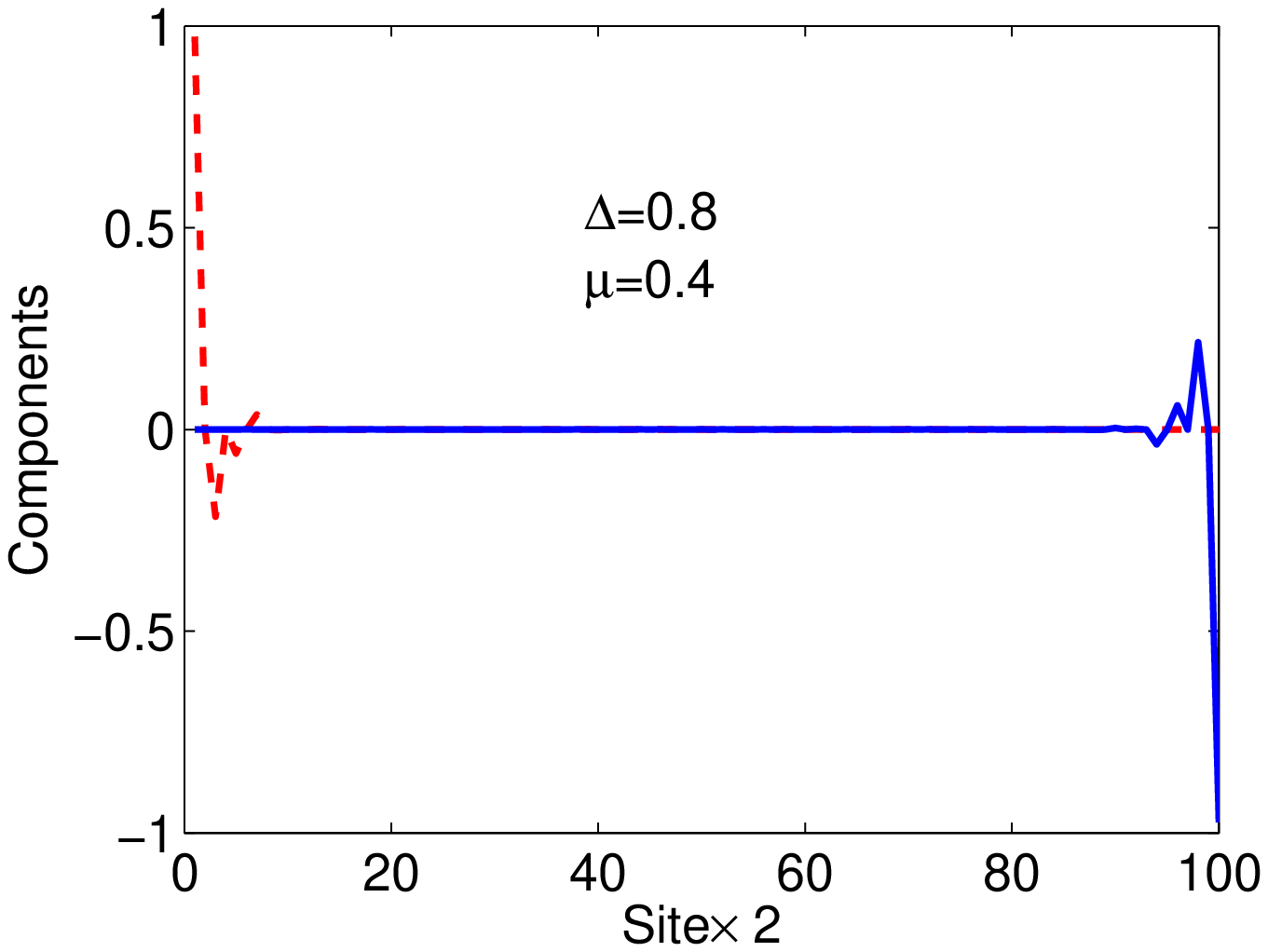}
\put(20,80){\text{(a)}}
\end{overpic}%
\begin{overpic}[width=2.8cm,height=2.8cm]{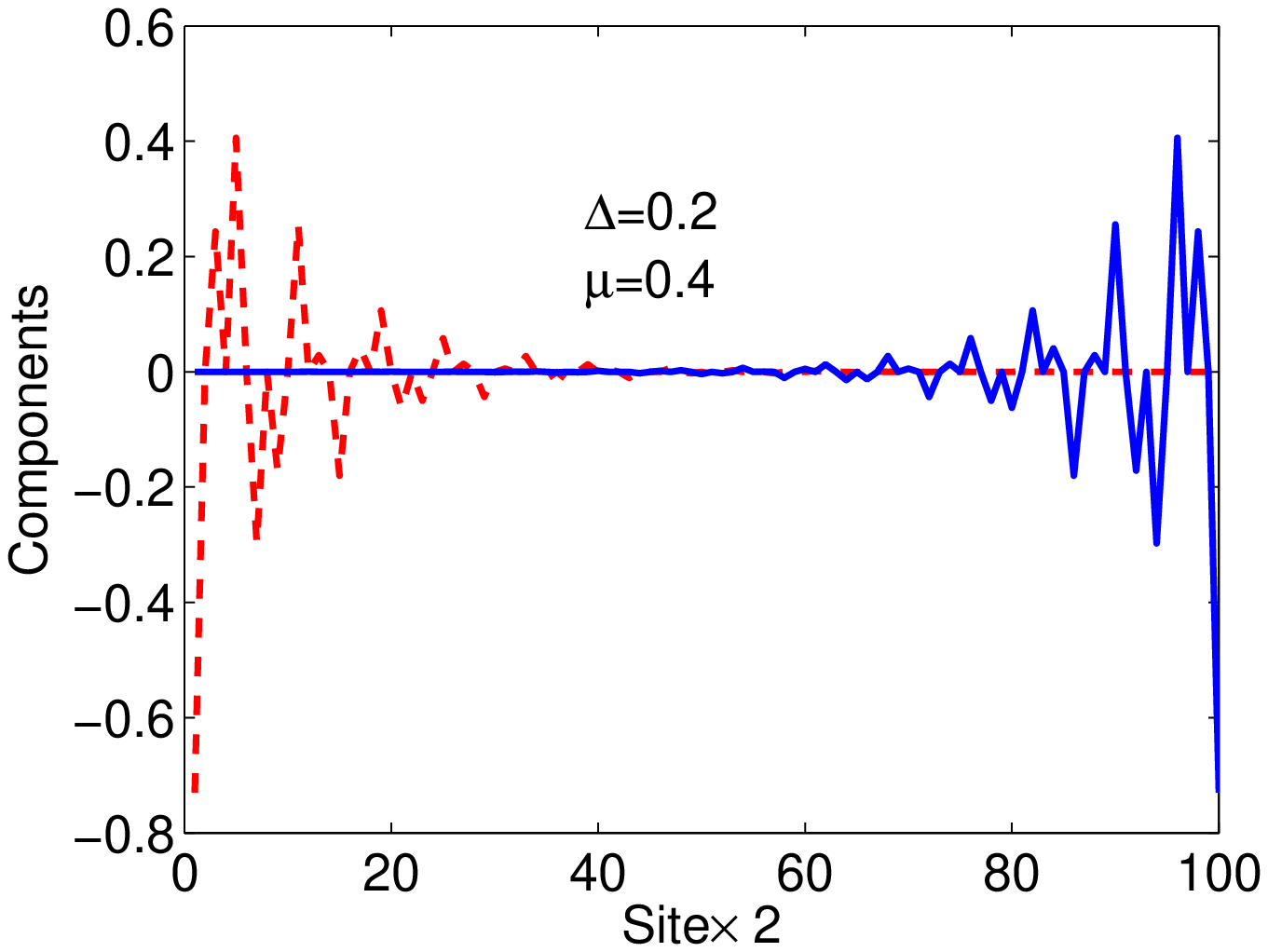}
\put(20,80){\text{(b)}}
\end{overpic}
\begin{overpic}[width=2.8cm,height=2.8cm]{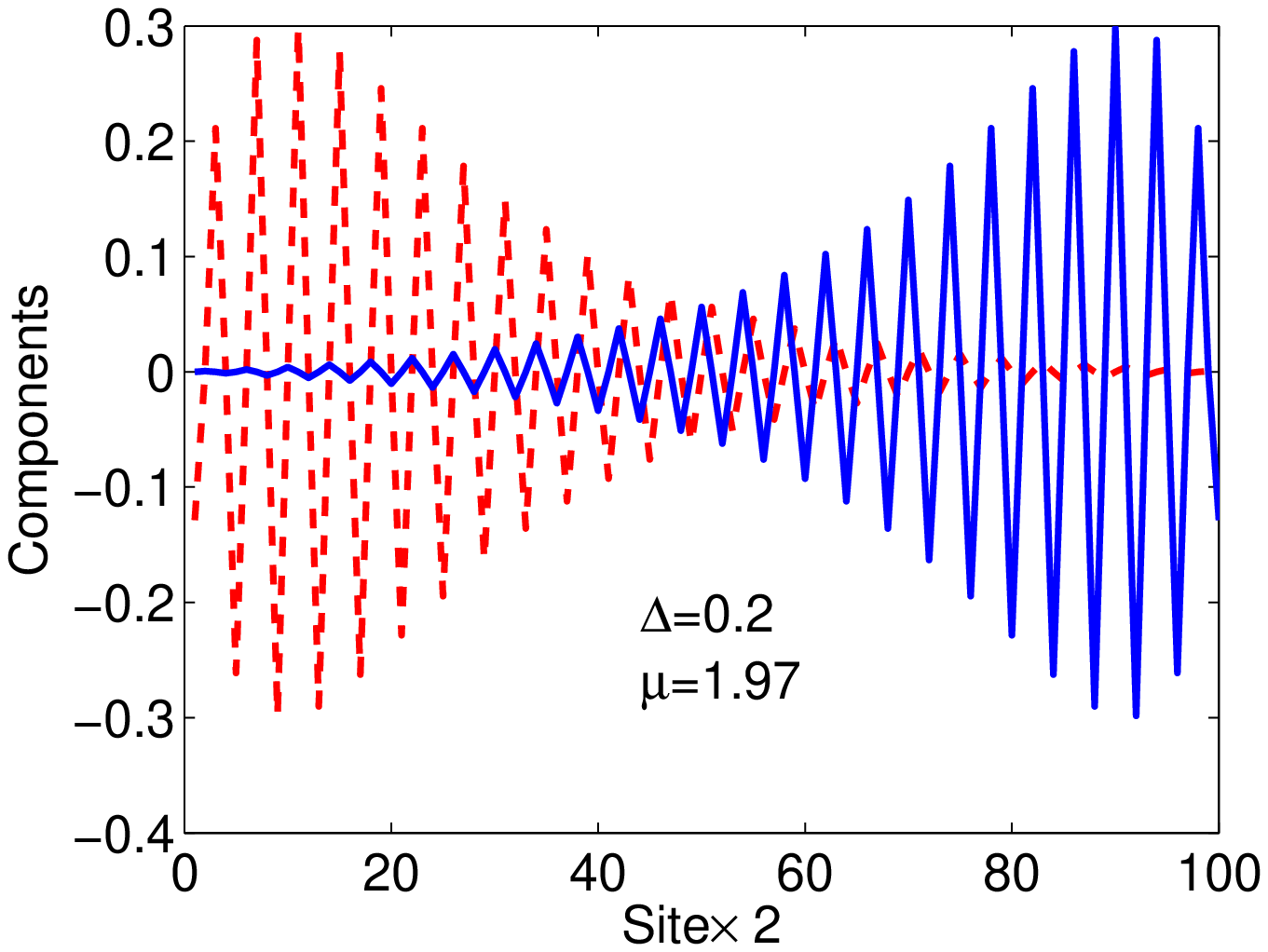}
\put(40,80){\text{(c)}}
\end{overpic}%
  \caption{Components on $\gamma_i$ of zero modes of a 50-site chain with different parameters (a) $t=1, \Delta=0.8, \mu=0.4$, (b) $t=1, \Delta=0.2, \mu=0.4$ and (c) $t=1, \Delta=0.2, \mu=1.97$. The solid (blue) and dashed (red) lines correspond to the two zero modes $\tilde\gamma_1$ and $\tilde\gamma_{2}$, respectively.  }
  \label{zeromode}
\end{figure}

The zero modes in a general case are calculated by a technique of Schur's decomposition. First the parameters are set in the domain $|\mu|<2t$ for a 50-site chain. Two zero modes appear with approximate zero energy $(<0.002)$. The components on Majorana operators $\gamma_i$ of the zero modes are plotted in Fig.\ref{zeromode}. As $|\Delta|$ deviates from $t$ the zero modes disperse gradually  away from the two ends but decay exponentially as expected. When $\mu$ approaches to $2t$, the phase transition point, the two zero modes overlap more and more and indeed disappear finally.

\begin{figure}
  \centering
\begin{overpic}[width=2.8cm,height=2.8cm]{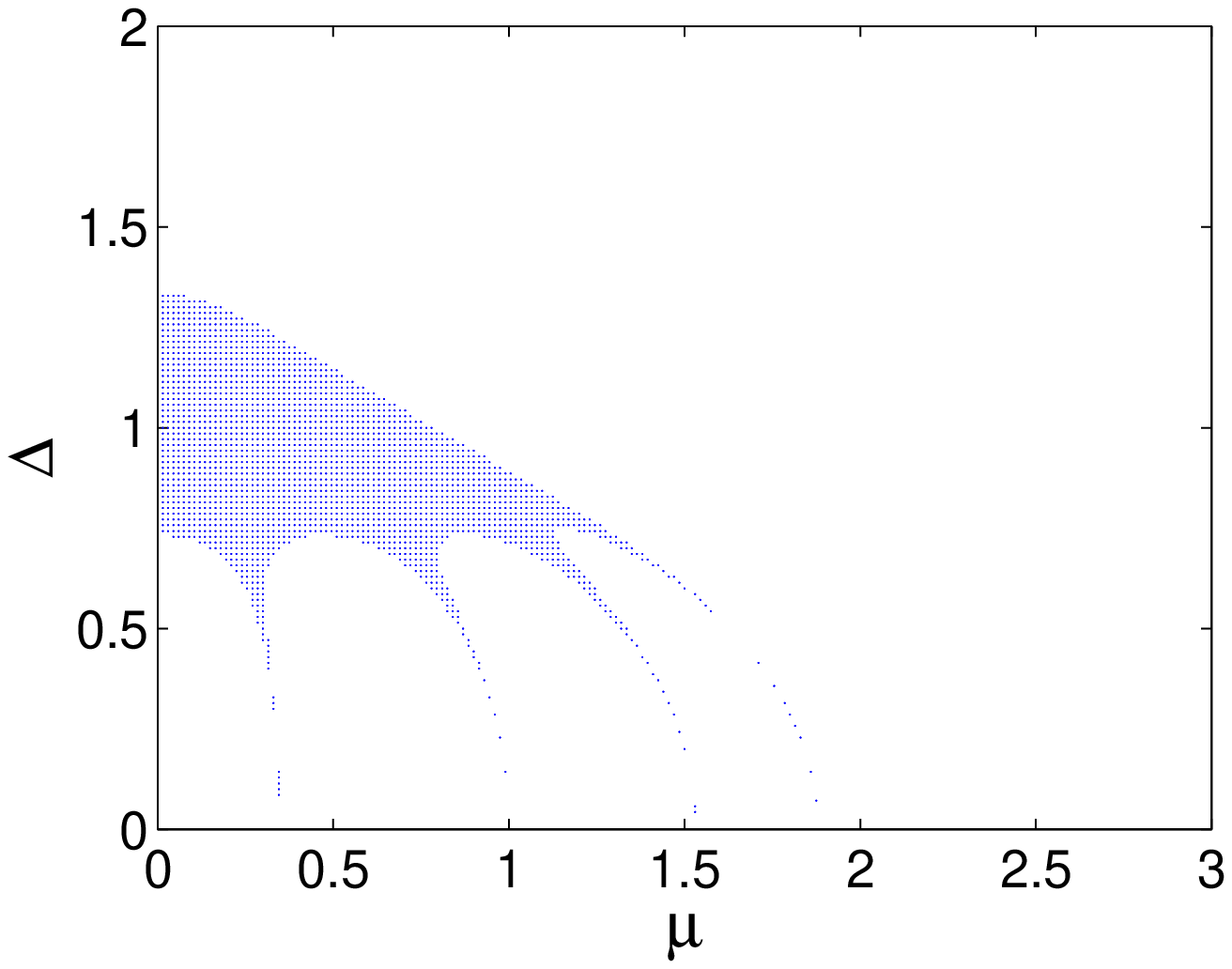}
\put(70,80){\text{(a)}}
\end{overpic}%
\begin{overpic}[width=2.8cm,height=2.8cm]{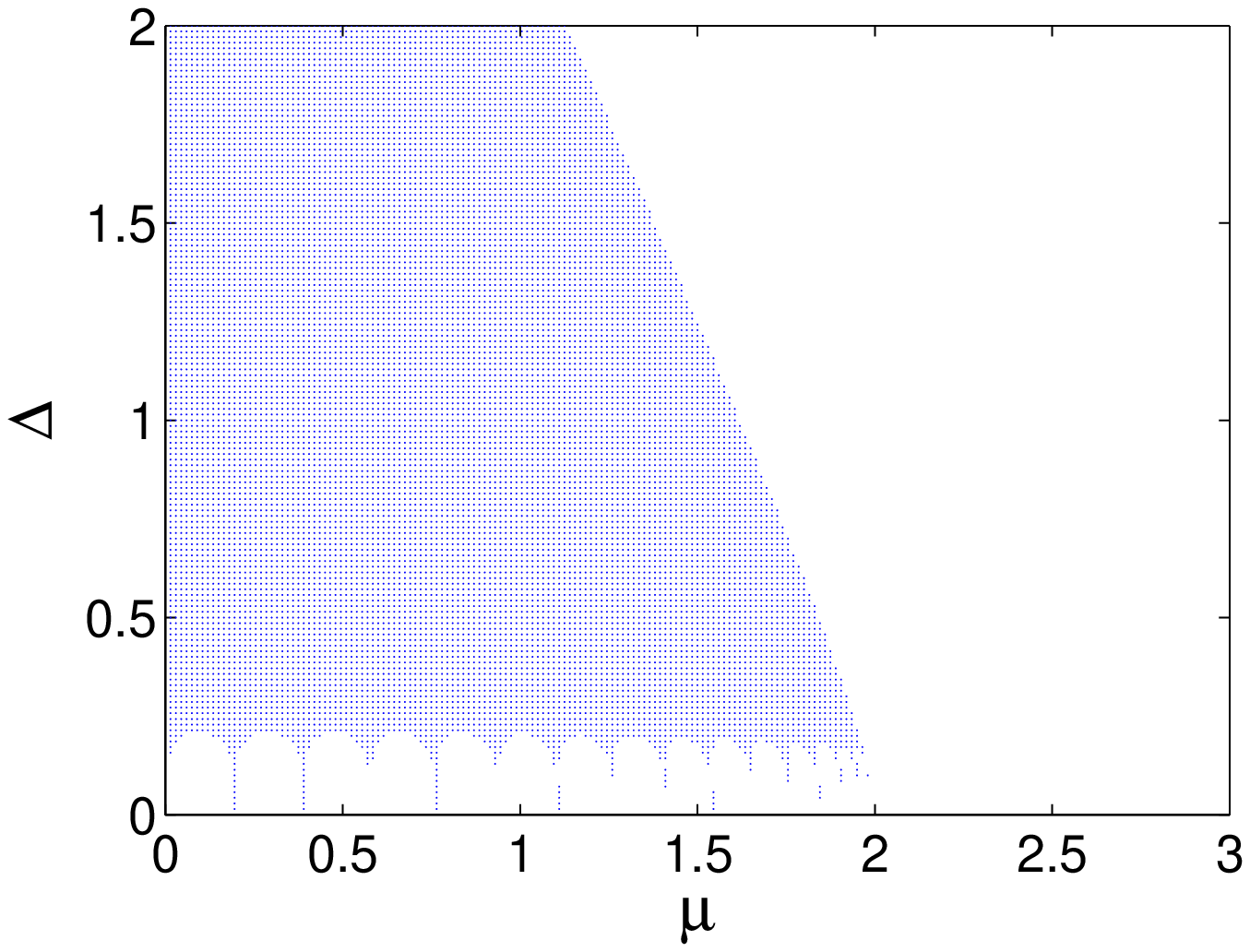}
\put(70,80){\text{(b)}}
\end{overpic}%
\begin{overpic}[width=2.8cm,height=2.8cm]{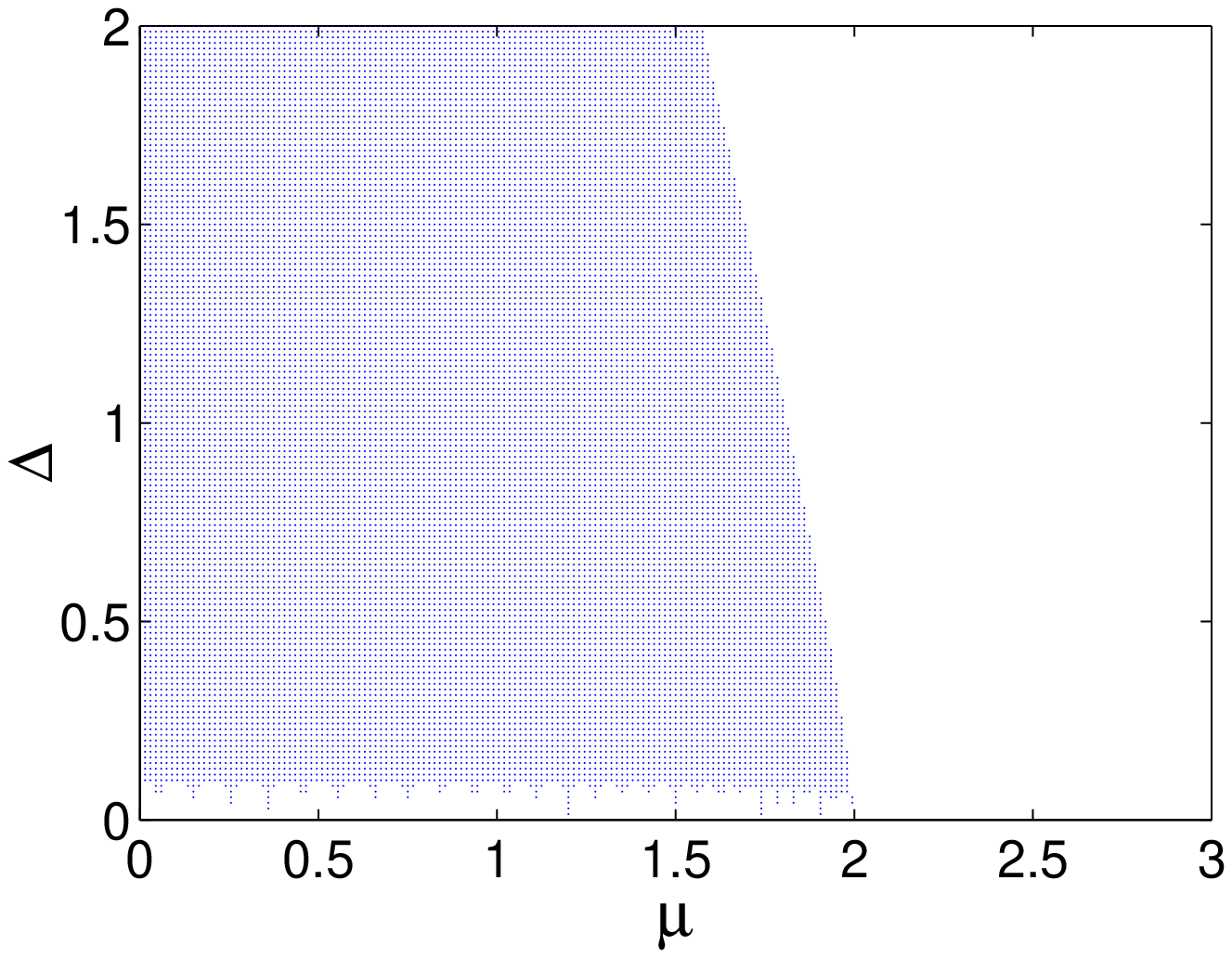}
\put(70,80){\text{(c)}}
\end{overpic}
  \caption{Phase diagrams of Majorana zero modes on chains with (a) 8 sites, (b) 31 sites  and (c) 60 sites. Shadow regions stand for the parameter domains of Majorana zero modes. 
  }
  \label{scan}
\end{figure}

To see the phase transition we plot the phase diagram of Majorana zero modes with respect to the superconductor gap $\Delta$ and the chemical potential $\mu$ in Fig.\ref{scan}.
It is interesting to see that the Majorana zero modes exist indeed below $\mu<2t$ but the boundary of the phase inclines to smaller $\mu$ for larger $\Delta$. The Majorana zero modes of a short chain have only a small parameter domain, deviating significantly from the predicted phase transition point $|\mu|=2t$. In addition, the Majorana zero modes have a threshold value for the superconducting gap. This indicates that a superconducting gap is necessary for the Majorana zero modes.


\section{Parities and robustness}

How the ground state looks like on the chain is an interesting issue. It is seen from the canonical form \eqref{cano} of the Hamiltonian that the ground state $|\psi\rangle$ satisfies $\langle\psi|\tilde{a}_m^{\dag} \tilde{a}_m|\psi\rangle =0 $ for $m\geq 2$ as $\epsilon_m>0$ but $\langle\psi|\tilde{a}_1^{\dag} \tilde{a}_1|\psi\rangle = 0$ or $1$ as  $ \epsilon_1=0$ for the Majorana zero modes. These two cases lead to two-fold degenerate ground states. In the occupation number representation  basis vectors can be expressed as
\begin{align}|\tilde n_N\tilde n_{N-1}...\tilde n_1\rangle=\prod_j^N (\tilde{a}_j^{\dag})^{\tilde n_j}|\tilde 0\rangle \end{align}
 where $|\tilde 0\rangle$ is one of the two degenerate ground states without any quasiparticles, i.e., $\tilde a_j|\tilde 0\rangle=0, j = 1,2,...,N$. Obviously, $|\tilde 0\rangle $ can be given by
\begin{align}
|\tilde 0\rangle &= C\tilde a_1 \tilde a_2 ...\tilde a_N|\text{vac}\rangle
\end{align}
where $C$ is a normalization factor and $|\text{vac}\rangle $ is the vacuum state. The other ground state is $|\tilde 1\rangle=\tilde{a}_1^{\dag}|\tilde 0\rangle$ which has a quasiparticle with zero energy. 

 One defines a parity operator in the following form
\begin{align}
\hat P &= \prod_{j=1}^N(-i\gamma_{2j-1}\gamma_{2j})=\prod_{j=1}^N(-2a_j^{\dag} a_j+1)
\end{align}
Since $\hat P^2=1$, $\hat P$ has two eigenvalues $\pm 1$, called even and odd parities, respectively. It can be easily verified that $[\hat P, H]=0$. This indicates that a non-degenerate eigenstate of $H$ must have a determinant parity, but degenerate eigenstates can be re-composed to have  parities. Most important is that an even(odd) parity corresponds to an superposition of states with even (odd) numbers of electrons.

The parity of the ground state $|\tilde 0\rangle$ can be worked out in the following way
\begin{align}
\hat P|\tilde 0\rangle &= (-1)^N C\tilde a_1 \tilde a_2 ...\tilde a_N\hat P|\text{vac}\rangle\nonumber \\&=(-1)^N|\tilde 0\rangle
\end{align}
where $\hat P|\text{vac}\rangle=|\text{vac}\rangle$ . Similarly, the other ground state has \begin{align}\hat P|\tilde 1\rangle=(-1)^{N+1}|\tilde 1\rangle\end{align}
Therefore, the two degenerate ground states, $|\tilde 0\rangle$ and $|\tilde 1\rangle$, have opposite parities, $(-1)^N$ and $(-1)^{N+1}$, and contain even (odd) and odd (even) numbers of electrons, respectively, for a chain with an even (odd) number of sites. A general ground state is a superposition of these two degenerate states: $\alpha|\tilde 0 \rangle + \beta|\tilde 1\rangle$, a parity mixing state.

\begin{figure}
\centering
\begin{overpic}[width=3.5cm,height=3.5cm]{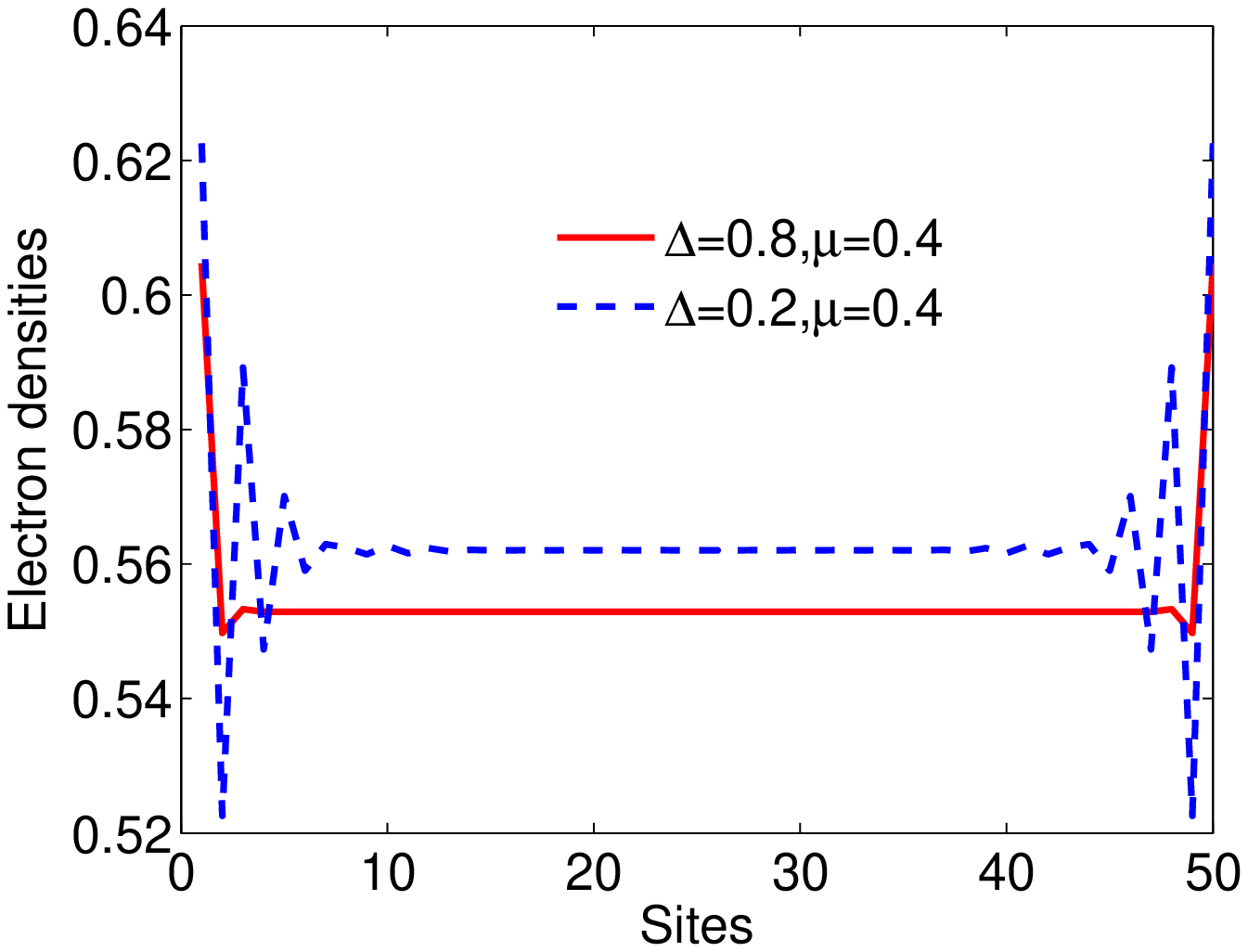}
\put(20,80){\text{(a)}}
\end{overpic}
\begin{overpic}[width=3.5cm,height=3.5cm]{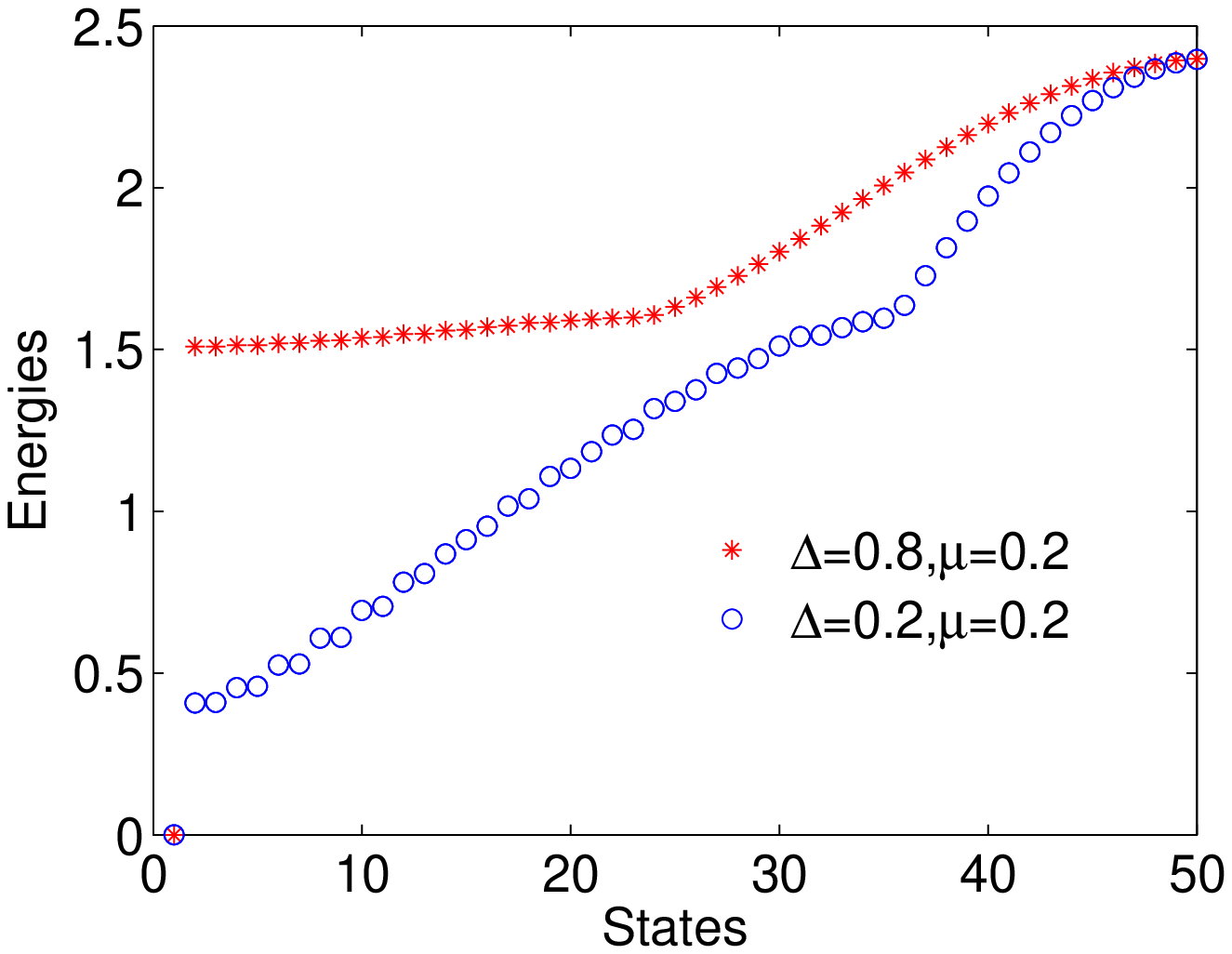}
\put(20,80){\text{(b)}}
\end{overpic}
\caption{(a) Electron density in the ground states and (b) single particle energies $\epsilon_m$ on a 50-site chain with parameters as shown.}
\label{ed}
\end{figure}

How do the electrons in the ground states distribute on the chain? The electron density is given by the expectation value $\langle a^\dagger_ja_j\rangle$ in the ground states. As shown in Fig.\ref{ed}(a) the electron density drops sharply about 10\% at both ends at $\Delta=0.8, \mu=0.4$ and oscillating around the ends at $\Delta=0.2, \mu=0.4$ on a 50-site chain, in accordance with  the Majorana zero modes shown in Fig.\ref{zeromode}(a,b). This provides a possibility to measure the Majorana zero modes. Electron densities in both ground states, however, are indistinguishable. The single particle energies are shown in Fig.\ref{ed}(b). There is a big gap between the zero mode energy and the none zero modes. This gap protects the Majorana zero modes to be robust for weak disturbance.

\begin{figure}
  \centering
  \begin{overpic}[width=1.7in]{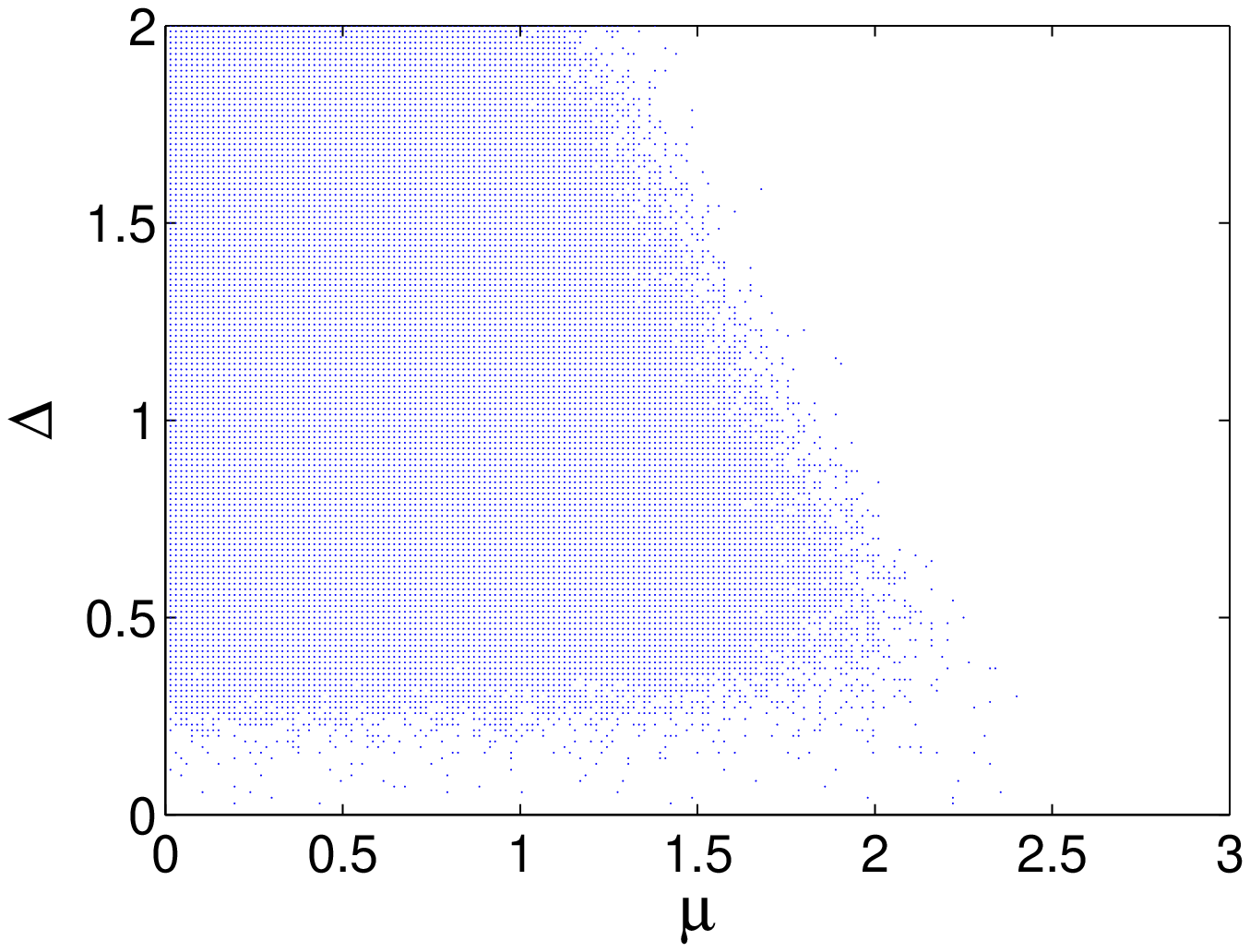}
\put(80,60){\text{(a)}}
\end{overpic}%
\begin{overpic}[width=1.7in]{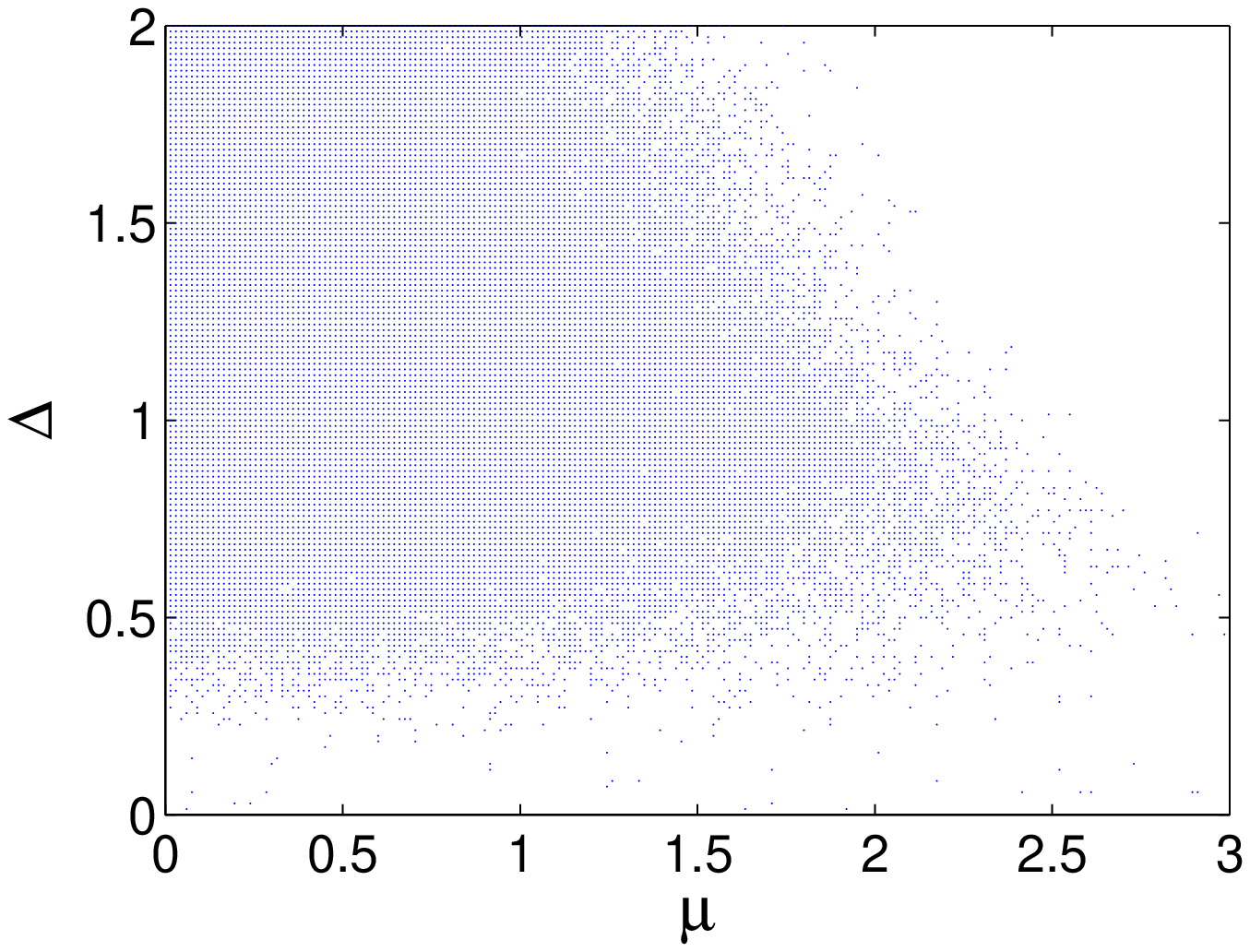}
\put(80,60){\text{(b)}}
\end{overpic}
 \caption{ Majorana zero modes (blue points) of a 31-site chain with noise (a) $V_0=1$ and (b) $V_0=2$.}\label{noise}
\end{figure}

How robust the Majorana zero modes are under the disturbance of noise is an essential issue for quantum computing. To simulate the disturbance of noise
we add to Hamiltonian \eqref{H} a local noise $H_n = \sum_j V_j a^\dagger_j a_j$ where the noise energies $V_j$ are simulated by $V_j = 2V_0 (R_j-1/2)$ with a series of random numbers $R_j\in (0,1)$. The phase diagrams after adding this noise are shown in Fig.\ref{noise}. It is seen that the boundary of the Majorana zero modes becomes more and more indistinct and diffuses into the forbidden region as increasing the intensity $V_0$,but the region with smaller chemical potentials is still robust under the disturbance.

\section{quantum computation}

Majorana zero modes distribute near the ends but do not couple to the quantum dot directly.
A real coupling occurs between the Majorana fermions at the end and the quantum dot.
Hence here we consider  the following Hamiltonian

\begin{align}
H &= H_0 + (V-\mu) (D^\dag D-\frac{1}{2}) + H_1 \\
H_1&=\frac{i}{2}(t+|\Delta|)\gamma_{2N}\Gamma_{1}+\frac{i}{2}(-t +|\Delta|)\gamma_{2N-1}\Gamma_{2}
\end{align}
where $\Gamma_{1}$ and $\Gamma_{2}$ denote Majorana fermion operators of the quantum dot at site $N+1$, $D^\dag=(\Gamma_{1} -i\Gamma_{2})/2$ denotes the creation of electrons on the quantum dot and $V$ is the bias voltage on the quantum dot. The operators $\gamma_{2N-1}$ and $\gamma_{2N}$ in the above Hamiltonian are given by
\begin{align}
\gamma_{J}=\sum_{j=1}^N(T^*_{J,j}\tilde a_j + T_{J,j}\tilde a^\dag_j), J =2N-1, 2N
\end{align}
where $T_{J,j}=W^T_{J,2j-1}+ iW^T_{J, 2j}$. They change the electron number thus the parity of the ground state. At high bias voltage this coupling results in nonzero excitations but it approaches to Karsten's result at small bias voltages near the ground states\cite{Karsten}.

\begin{figure}
  \centering
  \begin{overpic}[width=2.8cm]{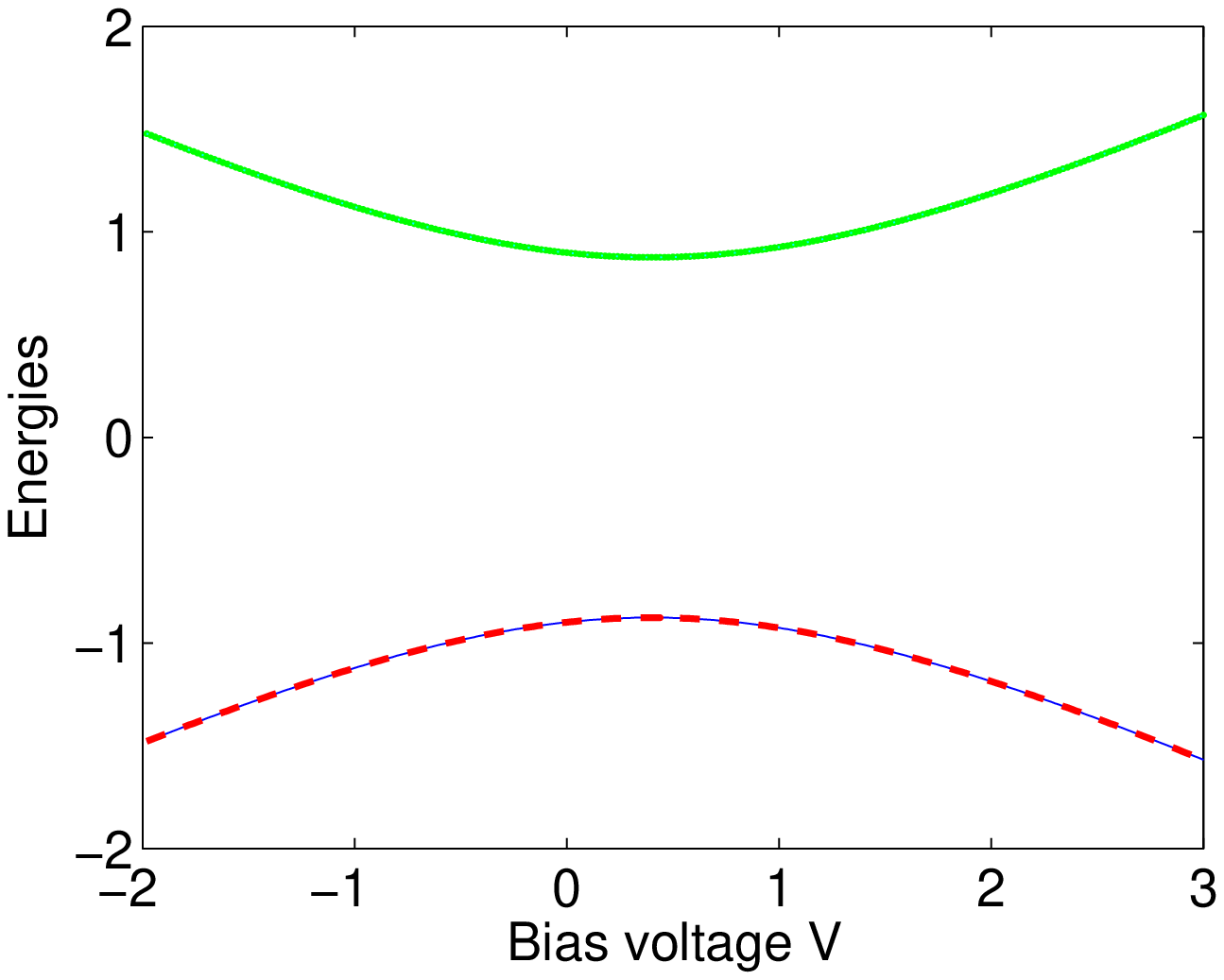}
\put(20,60){\text{(a)}}
\end{overpic}%
  \begin{overpic}[width=2.8cm]{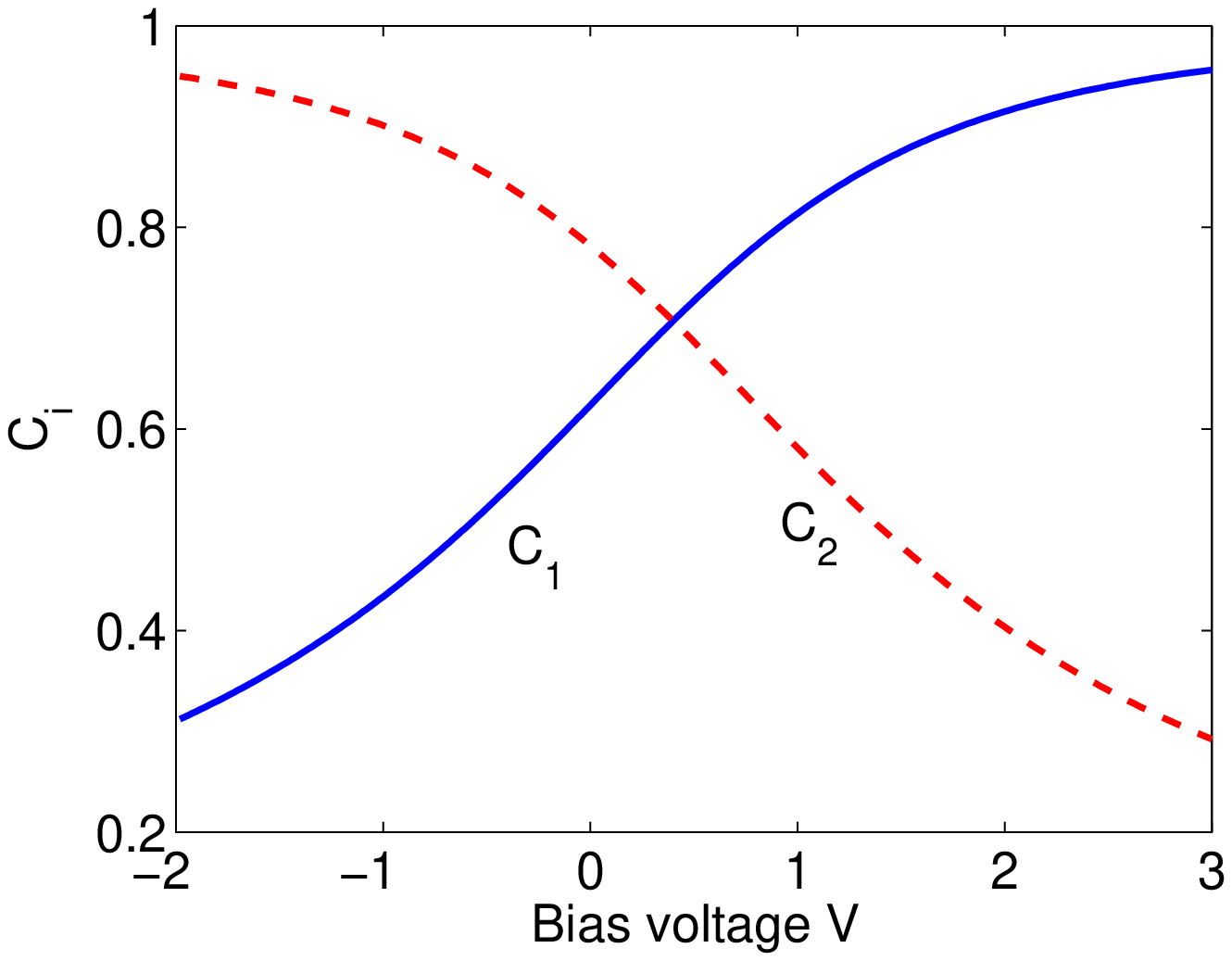}
\put(20,60){\text{(b)}}
\end{overpic}%
  \begin{overpic}[width=2.8cm]{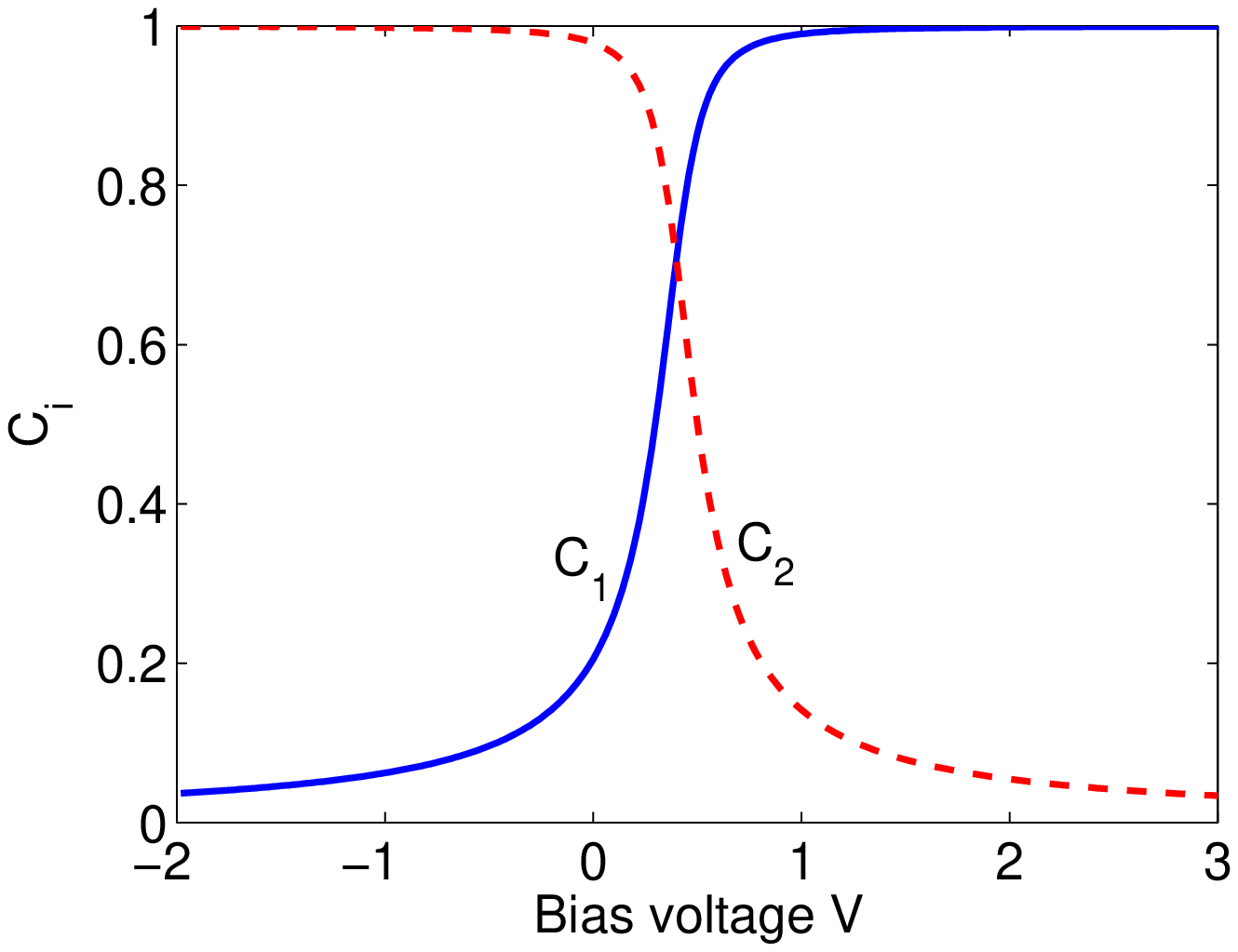}
\put(20,60){\text{(c)}}
\end{overpic}%
 \caption{(a) Energies of a quantum dot coupled to a 50-site chain with parameters $\Delta=0.8, \mu=0.4$; (b)Amplitudes $C_i$ in eq.\eqref{final} with the original coupling constant $S$;(c)Amplitudes $C_i$  with a 10-time reduced coupling constant $0.1S$. }
 \label{ci}
\end{figure}

We consider this coupling as a first-order perturbation on four near states $|\tilde n m\rangle =(\tilde a^\dag)^{\tilde n} (D^\dag)^m|\tilde0 0\rangle, \tilde n,m =0,1$ with unperturbed energies $-(V-\mu)/2,-(V-\mu)/2,(V-\mu)/2,(V-\mu)/2$.
Using the theory of the first-order perturbation we obtain two energies
\begin{align}E =\pm\sqrt{(V-\mu)^2/4 + |S|^2}\end{align}
where $S=-{1\over 2}(-t+|\Delta|)T_{2N-1,1}+{i\over 2}(t+|\Delta|)T_{2N,1}$ is the coupling constant. A superpostion of the two degenerate states  corresponding to the lower energy gives
\begin{align}
|\psi\rangle = C_1 (\alpha|\tilde 0 0\rangle + \beta|\tilde 10\rangle) + C_2(\alpha|\tilde 11\rangle + \beta|\tilde 01\rangle)\label{final}
\end{align}
where amplitudes $C_1={\rho/ \sqrt{ 1+|\rho|^2}}$,$C_2={1/ \sqrt{ 1+|\rho|^2}}$ with $\rho = -[(V-\mu)/2+\sqrt{(V-\mu)^2/4 + |S|^2}]/S$. This result is in coincidence with Karsten's result\cite{Karsten}. Through adiabatically tuning the bias voltage $V$ the above state can be driven from $|i\rangle=\alpha|\tilde 0 0\rangle + \beta|\tilde 10\rangle$ to $\tilde\gamma_1|i\rangle=\alpha|\tilde 11\rangle + \beta|\tilde 01\rangle$ continuously, where the parity of the chain has been reversed. This process, however, is limited by the gap above the ground states as shown in Fig.\ref{ed}(b). Since nonzero excitations may occur above the gap the initial state may be missing  in the tuning process. The energies of a 50-site chain coupled to a quantum dot and its amplitudes $C_i$ are plotted in Fig.\ref{ci} for different bias voltages. The bias voltage has been limited in the region so that the energy does not exceed the gap of the chain. It is seen from Fig.\ref{ci}(b) that a reversal of $C_1$ and $C_2$ is truly taking place but they do not have a clear 0 to 1 reversal. Reducing the coupling constant there will be a clearer reversal as shown in Fig.\ref{ci}(c) for a coupling constant $0.1S$. Therefore, a parity reversion can be realized in a weak coupling between a chain and quantum dot. 


\section{Summary and conclusion}

In this work numerical calculations for Majorana zero modes on a one-dimensional chain are performed using the technique of block diagonalization in Schur's decomposition for a general parameter setting. It is found that Majorana zero modes occur near the ends of the chain and decay exponentially away from the ends. The phase diagrams show that Majorana zero modes of a long-enough chain indeed have a parameter domain of $2t>|\mu|$ as predicted from the bulk property of the chain, but a short chain has a much smaller parameter domain than the prediction. Through a simulation Majorana zero modes are found to be robust under the disturbance of noise. Finally the reversion of the parity of the ground states is studied by applying a bias voltage on an end of the chain. For a weak coupling between a chain and a quantum dot the parity of the ground states can be reversed through adiabatically tuning the bias voltage.

\end{document}